\documentclass[epj]{svjour}
\usepackage{graphicx}
\usepackage{epsfig}
\usepackage{amssymb}

\title{Recent high-$p_{T}$ results from STAR}
\author{M. van Leeuwen \inst{1} {\it for the STAR collaboration}}
\institute{Universiteit Utrecht, PO Box 80000, Utrecht, Netherlands \email{m.vanleeuwen1@uu.nl}}

\newcommand{\pt}{\ensuremath{p_{T}}}
\newcommand{\pttrig}{\ensuremath{p_{T}^{trig}}}
\newcommand{\ptassoc}{\ensuremath{p_{T}^{assoc}}}
\newcommand{\dphi}{\ensuremath{\Delta\phi}}
\newcommand{\deta}{\ensuremath{\Delta\eta}}

\abstract{
We present selected recent results of multi-hadron correlation
measurements in azimuth and pseudorapidity at intermediate and high
\pt{} in Au+Au collisions at $\sqrt{s_{NN}}=200$ GeV, from the
STAR experiment at RHIC. At intermediate \pt, measurements are
presented that attempt to determine the origin of the associated
near-side (small \dphi) yield
at large pseudo-rapidity difference \deta{} that is found to be
present in heavy ion collisions. In addition, results are reported on
new multi-hadron correlation measures at high-\pt{} that use di-hadron
triggers and multi-hadron cluster triggers with the goal to 
constrain the underlying jet kinematics better than in the existing
measurements of inclusive spectra and di-hadron correlations.
}

\begin{document}

\maketitle

The goal of research with high-energy nuclear collisions at the
Relativistic Heavy Ion Collider (RHIC) is to study
bulk matter systems where the strong nuclear force as described by
Quantum Chromo Dynamics (QCD) is the dominant interaction. In
particular, it is expected that there is a phase transition of bulk
QCD matter to a deconfined state at high temperature. 

In heavy ion collisions, products of initial state hard scatterings,
such as high-\pt{} hadrons, photons and heavy mesons, can be used to
probe the soft matter generated in the collision. Initial state
production of high-\pt{} partons is relatively unaffected by the
presence of the soft medium, but the partons lose energy when
traversing the medium, dominantly due to gluon radiation. The goal of
high-\pt{} measurements is to study these interactions and to use them
to measure the density and temperature of the soft matter.

It has also been found at RHIC that at intermediate $\pt \approx 2-6$
GeV/$c$, the baryon/meson ratio is much larger in heavy ion collisions
than in proton-proton collisions \cite{Abelev:2006jr,Abelev:2007ra}. This could
imply that soft production mechanisms, {\it e.g.} hadron formation by
coalescence of quarks from thermal matter
\cite{Lin:2002rw,Hwa:2002tu,Fries:2003vb}, have a significant
contribution to baryon (and meson) production up to $\pt \sim 6$
GeV/$c$.

The underlying (di-)jet structure for particle production at
high-\pt{} can be probed using azimuthal di-hadron correlations, which measure
the associated particle distribution in azimuth with respect to some higher
\pt{} `trigger' particle \cite{Adams:2006yt}. This technique has been
applied with trigger hadrons from low to high \pt. In the
following, we will first discuss di-hadron measurements at
intermediate \pt, where the soft matter from the medium seems to contribute to
correlation structures. The last part of these proceedings deals with
high-\pt{} measurements, where jet-quenching followed by
jet-fragmentation in the vacuum seems to dominate.

\section{Intermediate \pt: near-side ridge}
\begin{figure}
\epsfig{file=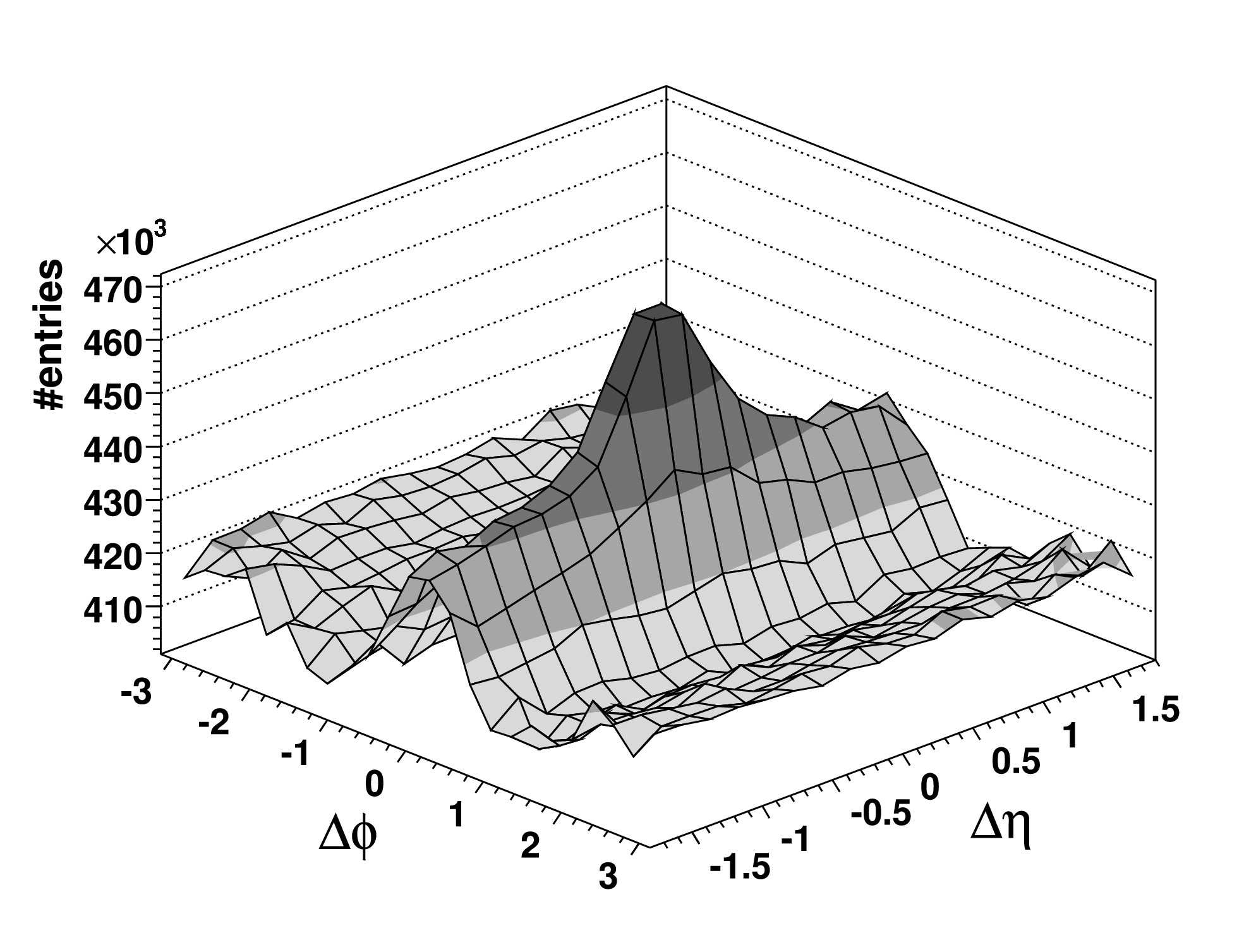,width=0.45\textwidth}
\put(-95,120){\sf STAR Preliminary}

\epsfig{file=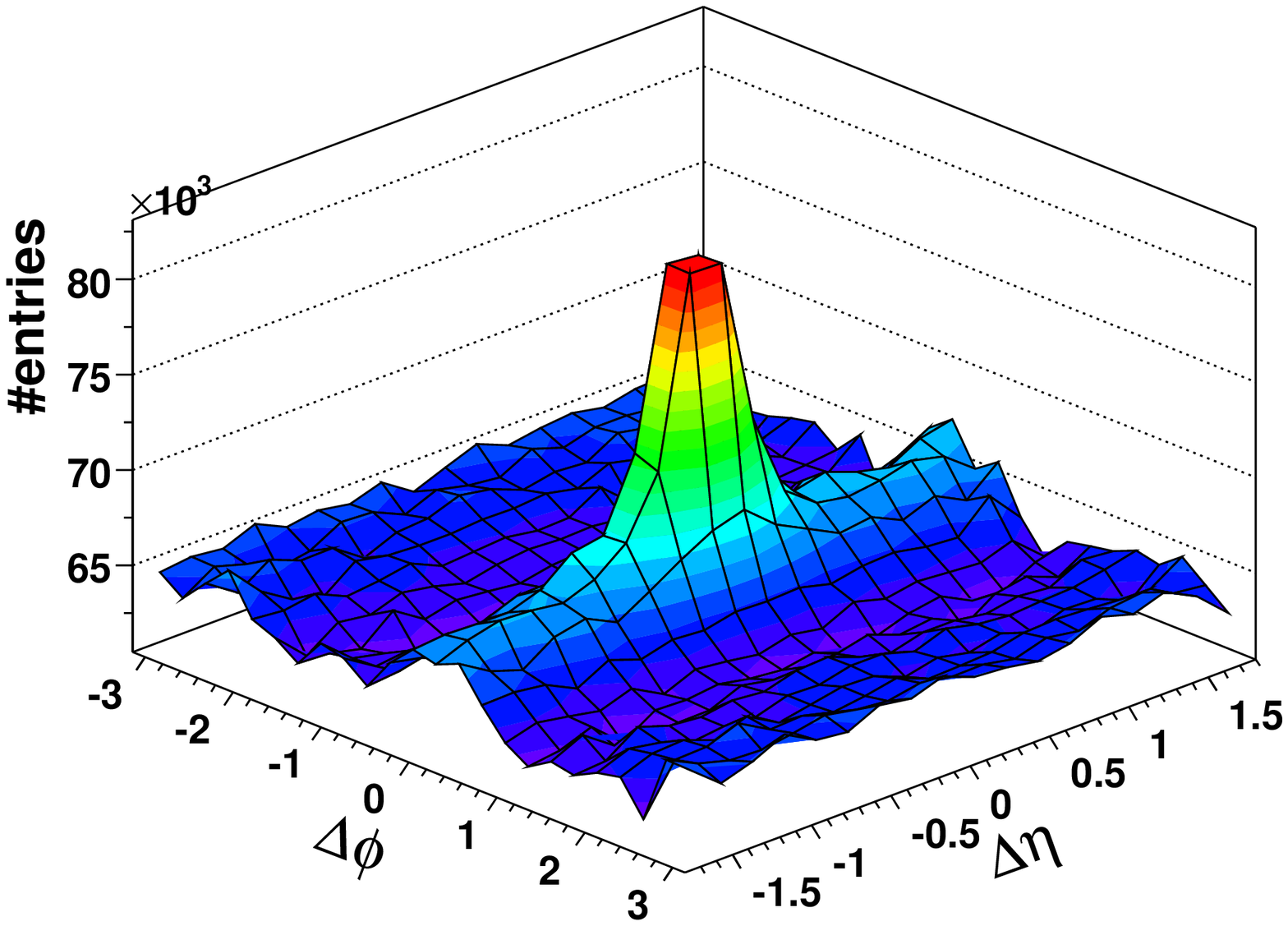,width=0.49\textwidth}
\put(-110,120){\sf STAR Preliminary}
\caption{\label{fig:ridge_2D}(Color on-line) Preliminary associated
particle distributions in \deta{} and \dphi{} with respect to the
trigger hadron for associated particles with $2~ \mathrm{GeV}/c <
\ptassoc < \pttrig$ in 0-12\% central Au+Au collisions. Two different
trigger \pt{} selections are shown: $3 < \pttrig < 4$ GeV/$c$ (upper
panel) and $4 < \pttrig < 6$ GeV/$c$ (lower panel). No background was
subtracted.}
\end{figure}
One of the striking results in di-hadron correlations at intermediate
\pt{} at RHIC is that near-side (small \dphi) associated yield is
observed at large \deta{} with respect to the trigger in heavy ion
collisions. This is illustrated in Fig. \ref{fig:ridge_2D} which shows
associated particle distributions in pseudo-rapidity difference
\deta{} and azimuthal angle difference \dphi, with respect to the
trigger hadron direction for associated particles with $2~
\mathrm{GeV}/c < \ptassoc < \pttrig$. The upper panel shows the
results for trigger hadrons $3 < \pttrig < 4$ GeV/$c$ and the lower
panel for $4 < \pttrig < 6$ GeV/$c$. The associated particle
distribution shows a clear peak at $(\deta,\dphi) \approx (0,0)$ as
expected from jet fragmentation. Additional associated yield is seen
at large $\deta \gtrsim 1$, which is unique to heavy ion
collisions. Within the experimental acceptance, the additional yield
is approximately independent of \deta, and therefore referred to as
the {\it ridge}. For higher \pttrig{} (lower panel of
Fig. \ref{fig:ridge_2D}), the ridge yield is less prominent, because
the yield in the jet-like peak increases with \pttrig{}. Quantitative
analysis of the yields shows that ridge yield is independent of
\pttrig{} within the statistical and systematic uncertainties
\cite{Putschke:2007mi}.

Various production mechanisms for the ridge have been proposed in the
literature. Here we would like to classify these mechanisms into two
broad categories, namely 'jet-like' and 'bulk-like'. Jet-like production
mechanisms are based on the idea that the ridge is formed mainly from jet fragments,
which couple to the longitudinal flow
\cite{Armesto:2004pt,Majumder:2006wi,Romatschke:2006bb}. Bulk-like
production mechanisms, on the other hand, start from the assumption
that the passage
of the jet through the longitudinally extended bulk matter locally
increases the yield, for example by increasing the mean-\pt{} through
collisions \cite{Wong:2007pz}, or heating of the medium \cite{Chiu:2005ad}. A
variant on these bulk-like mechanisms is that the
enhancement does not need to be due to interactions between the jet
and the medium, but could be the result of radial flow combined with trigger
bias \cite{Voloshin:2004th,Pruneau:2007ua}.

Here, we would like to highlight two recent results from STAR that may
shed further light on the origin of the ridge.

\begin{figure}
\epsfig{file=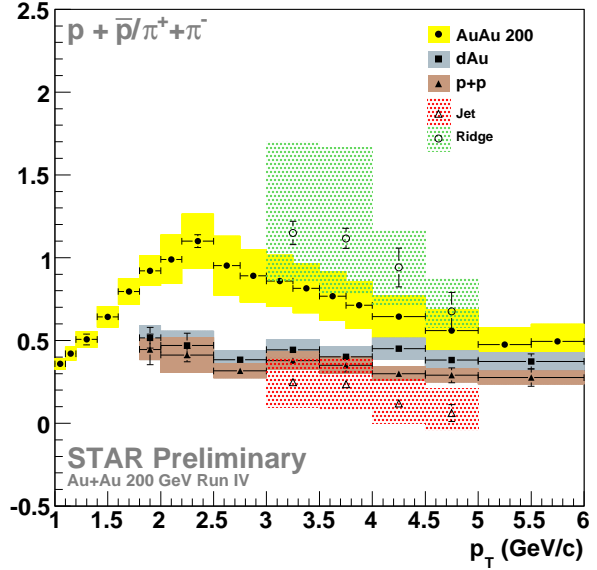,width=\columnwidth}
\caption{\label{fig:ppi_ridge}(Color on-line) Preliminary p/$\pi$
  ratio in the ridge and jet-like peak and a function of \ptassoc{}
  for $4<\pttrig<6$ GeV/$c$ in 0-12\% central Au+Au collisions (open
  symbols). For comparison, inclusive p/$\pi$ ratios are shown with
  solid symbols.}
\end{figure}
Figure \ref{fig:ppi_ridge} shows the p/$\pi$ ratio in the ridge
and jet-like peak as a function of
\ptassoc{} for $4<\pttrig<6$ GeV/$c$. The inclusive p/$\pi$ ratios are
also shown for comparison. Clearly, the p/$\pi$ ratio in
the ridge is similar to the inclusive p/$\pi$ ratio in Au+Au events,
which is much larger than in p+p events. The
p/$\pi$ ratio in the jet-like peak is similar to the inclusive ratio
in p+p events. These results clearly suggest that the ridge is
formed from bulk matter and not from jet fragments.

\begin{figure*}
\centering
\epsfig{file=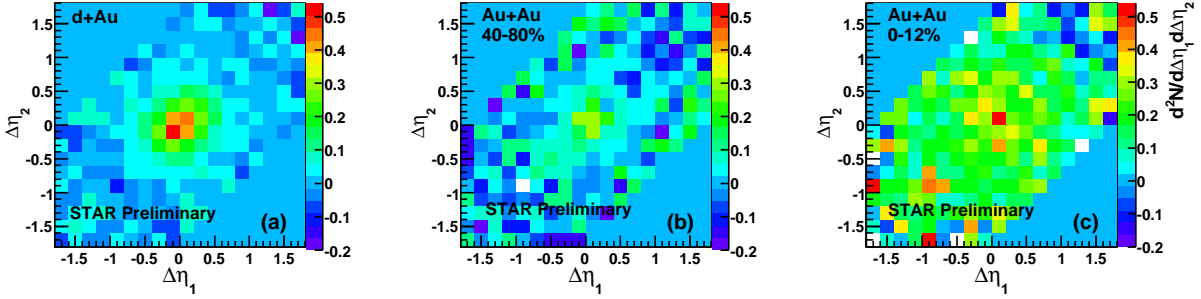,width=0.9\textwidth}
  \caption{\label{fig:detadeta}(Color on-line) Background-subtracted
associated hadron pair distribution as a function of $\deta_1$ and
$\deta_2$ the pseudo-rapidity difference between the trigger hadron and
the first and second associated hadron for particles with $1 <
\ptassoc < 3$ GeV/$c$ and $3 < \pttrig < 10$ GeV/$c$. The left panel
shows results for d+Au collisions, center and right panel are Au+Au at
40-80\% and 0-12\% most central.}
\end{figure*}
The second result that addresses the origin of the ridge yield is the
analysis of three-particle $\deta-\deta$ correlations. Figure
\ref{fig:detadeta} shows the distribution of associated hadron pairs
as a function of $\deta_1$ and $\deta_2$, the pseudo-rapidity difference
between the trigger hadron and the first and second associated hadron,
for particles with $1 < \ptassoc < 3$ GeV/$c$ and $3 < \pttrig < 10$
GeV/$c$. Combinatorial backgrounds have been subtracted (for details
on the procedure, see \cite{Netrakanti:2008jw}). This measurement is
sensitive to the event-by-event substructure of the ridge: an excess
along the diagonal would indicate that ridge particles tend to be
close together in $\eta$, which would be expected if the ridge hadrons
are fragmentation products of radiated gluons that are carried along
with the longitudinal flow. No such excess is observed, indicating
that (within the statistical reach of the analysis) the particles in
the ridge are distributed evenly in $\eta$ in every event. Another
remarkable feature in the figure is the absence of a horizontal and
vertical band (i.e. $\deta_1\approx 0$ or $\deta_2 \approx 0$), which
should arise from events where one associated particle is inside the
jet-like peak and the other is part of the ridge. Further work is
ongoing to determine how significant the absence of the various
structures is. Future RHIC runs will provide increased statistics
which will reduce the uncertainties and will make it possible to
perform this analysis
with larger $\pttrig$ and $\ptassoc$, where backgrounds are smaller.

\section{Path length dependence}
\begin{figure*}
  \epsfig{file=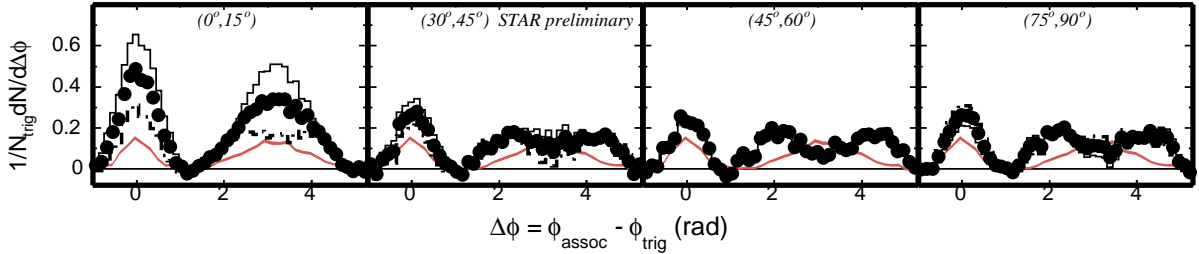,width=0.98\textwidth}
  \caption{\label{fig:dphi_react_plane} Preliminary background
subtracted distributions of associated hadrons with $1.0 < \ptassoc <
1.5$ GeV/$c$ relative to a trigger hadron $3.0 < \pttrig < 4.0$
GeV/$c$ in four different ranges for the angle $\phi_s$ between the
trigger hadron and the event plane for 20-60\% central Au+Au
collisions. The solid and dashed histogram represent the systematic
uncertainty due to the uncertainty in the strength of the elliptic
flow $v_2$. The (red) line shows the result from d+Au collisions for
reference.}
\end{figure*}
To vary the path length of high-\pt{} partons
through the medium, measurements are performed as a function of
the angle with respect to the event plane. Figure
\ref{fig:dphi_react_plane} shows the background subtracted
distributions  of associated hadrons with $1.0 < \ptassoc < 1.5$
GeV/$c$ relative to a trigger hadron $3.0 < \pttrig < 4.0$ GeV/$c$ in
four different ranges for the angle $\phi_s$ between the trigger hadron and the
event plane. On the near side, a decrease of the associated yield with
increasing $\phi_s$ is visible. The recoil peak shape changes from a
broad single peak at $\phi_s=0$ to a doubly-peaked structure for
larger angles. The systematic uncertainty on the background
subtraction due to the uncertainty in elliptic flow $v_2$ is indicated
by the dashed and solid histograms. For details of the analysis, see \cite{Feng:2008an}.

\begin{figure}
  \epsfig{file=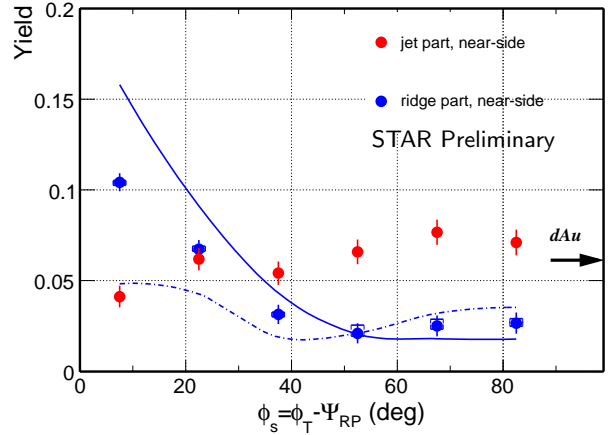,width=0.45\textwidth}
  \put(-90,110){\sf STAR Preliminary}
  \caption{\label{fig:ridge_phis} Preliminary results for the
    near-side yield as a function of the angle between the event
    plane and the trigger hadron ($3.0 < \pttrig < 4.0$ GeV/$c$), for
    associated hadrons $1.5 < \ptassoc < 2.0$ GeV/$c$ for 20-60\%
    central Au+Au collisions. Red circles show the jet-like yield, and
    blue circles show the ridge yield.}
\end{figure}

Figure \ref{fig:ridge_phis} shows the dependence of the near-side
yield on $\phi_s$, for the ridge and the jet-like peak separately. It
can be seen that the jet-like peak has almost no dependence on the
angle with respect to the event plane, while the ridge component is
significantly larger for small $\phi_s$, {\it i.e.} in-plane. This
implies that the ridge yield is larger for smaller path lengths, while
naively one would expect larger jet modifications for longer
path lengths. The observed trend might be due to a trigger bias effect,
for example because jets with large energy loss do not give rise to a
trigger particle. Another possible interpretation is that the ridge
effect may depend on the local flow velocity in the medium.

\begin{figure}
  \epsfig{file=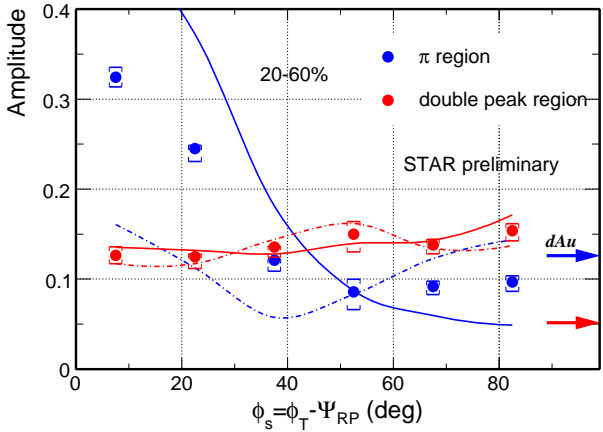,width=0.45\textwidth}
  \caption{\label{fig:recoil_phis} Preliminary results for the recoil
    yield as a function of the angle between the event plane and the
    trigger hadron ($3.0 < \pttrig < 4.0$ GeV/$c$), for associated
    hadrons $1.0 < \ptassoc < 1.5$ GeV/$c$ for 20-60\% central Au+Au
    collisions. Blue circles show the yield in $|\dphi -\pi|<0.42$
    (jet-region) and red circles show the yield at
    $0.78<|\dphi-\pi|<1.65$.}
\end{figure}
Figure \ref{fig:recoil_phis} shows the dependence of the recoil yield
in two angular regions around $\dphi=\pi$ and $\dphi\approx2/3\pi$
(`double peak region') on
$\phi_s$. The associated yield at $\dphi\approx2/3\pi$ shows no
significant dependence on $\phi_s$, while the yield at $\dphi=\pi$ is
largest in-plane ($\phi_s=0$) and then decreases. This indicates that
the change of the shape of the recoil distribution is mostly due to
changing yield at $\dphi=\pi$. The doubly-peaked away-side occurs when
the yield at $\dphi=\pi$ is smaller than in the `double peak
region'. It is also interesting to note that while the double peak
structure is associated with large path length, the strongest increase
of the yield at $\dphi=\pi$ with respect to d+Au collisions is seen
for the shortest path lengths, $\phi_s=0$.

An important systematic effect in the measurement of azimuthal
correlations as a function of the angle with respect to the event
plane, is that the jet-like azimuthal correlations may bias the event
plane. In order to reduce this effect, the event plane is
reconstructed using tracks with $\deta > 0.5$ with respect to the
trigger hadron. Some of the observed effects, however, could be due to
such an event plane bias, so further tests of this systematic effect
are still ongoing.

\section{High \pt{} multi-hadron correlations}
At high $\pt \gtrsim 6$ GeV/$c$, both the hadrochemistry, as measured
by the baryon/meson ratio \cite{Abelev:2006jr} and the jet-like peak
shapes in the di-hadron analysis \cite{Adams:2006yt} are very similar
in heavy ion collisions and p+p collisions, which suggests that at
high \pt, the dominant particle production mechanism in heavy ion
collisions is parton fragmentation.

In-medium parton energy loss leads to a suppression of high-\pt{} hadron
production, which can be modeled using perturbative QCD
techniques. Recently, progress has being made towards determining
medium properties by systematically confronting the data with several
models for energy loss
\cite{Dainese:2004te,Zhang:2007ja,Adare:2008cg}. However, inclusive
spectra and di-hadron correlations are rather insensitive to details
of the energy loss process, such as the probability distribution for
energy loss, because the initial parton energy is not measured
\cite{Renk:2007mv}. As a result, a number of energy loss models are
able to describe the data, despite significant differences in the
underlying energy loss distributions. To further address this, more
differential measurements are needed, preferable including a direct
measure of the initial parton energy.

Direct-photon jet measurements are especially promising in this regard
\cite{Wang:1996yh,Renk:2006qg}, since the photon energy is a direct
measure of the initial parton energy. Direct photon measurements,
however, have the difficulty that the event-samples are smaller than
for hadronic signatures and that the backgrounds are large. Another
approach is to perform jet reconstruction in heavy ion collisions,
where the jet energy is a measure of the initial parton
energy. Results from both types of measurement will be discussed in
other contributions to this volume \cite{hamed,heinz,putschke,salur}.

Here, we would like to present two analyses which provide more
differential information that the inclusive spectra and di-hadrons,
but do not provide a direct measurement of the initial parton energies.

\subsection{Di-hadron triggered correlations}
\begin{figure}
  \psfig{file=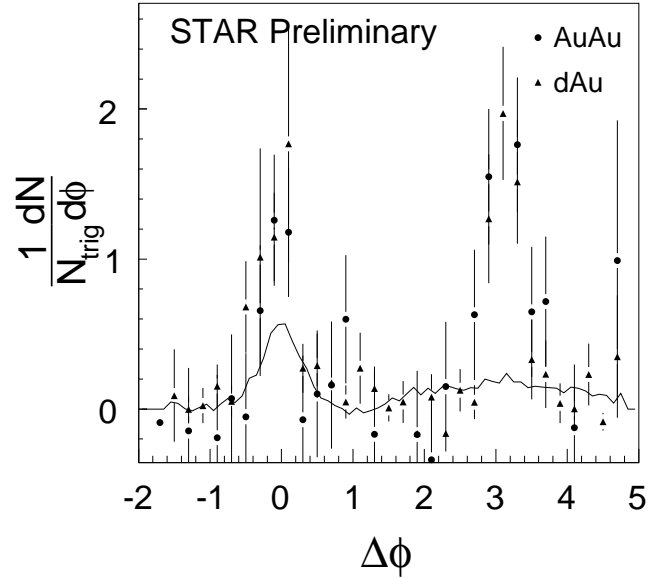,width=0.48\textwidth}
  \caption{\label{fig:dijet_trig}Preliminary associated hadron
  distribution with respect to a di-hadron trigger with back-to-back
  hadrons of $5< p_{T,1}^{trig}<10$ GeV/$c$ and $4 < p_{T,2}^{trig} <
  5$ GeV/$c$ in 0-12\% central Au+Au collisions. Associated hadrons
  have $1.5 < \ptassoc < 4$ GeV/$c$. The angular difference
  $\dphi=\phi_A-\phi_{T2}$ is between the associated hadron and the
  second trigger. Combinatorial backgrounds,
  including jet-background combinations, have been subtracted. The
  line shows the di-hadron distribution with $4 < \pttrig < 5$ GeV/$c$
  and $1.5 < \ptassoc < 4$ GeV/$c$ from central Au+Au collisions for
  comparison.}
\end{figure}
In the standard di-hadron measurement \cite{Adams:2006yt}, the
near-side trigger hadron \pt{} selects jets within a relatively broad
energy range. As a result, the \pt-cut on associated hadrons on the away-side
can impose an additional bias on the underlying jet
distribution. The analysis presented here aims to reduce this effect
by using a di-jet
trigger, {\it i.e.} selecting events with a back-to-back pair of
high-\pt{} hadrons. Associated hadrons can then be studied with
respect to this pair.

Figure \ref{fig:dijet_trig} presents a first result from STAR for this
type of analysis, using trigger
hadron pairs with $5 < p_{T,1}^{trig}<10$ GeV/$c$ and $4 < p_{T,2}^{trig} < 5$
GeV/$c$ and associated hadrons $1.5 < \ptassoc < 4$
GeV/$c$. The combinatorial backgrounds have been subtracted,
including jet-background cross-terms. For details, see
\cite{barannikova}. For comparison, the line shows the associated
hadron distribution for a single trigger particle with $4 < \pttrig <
5$  GeV/$c$ and the same \ptassoc{} selection as the di-hadron
trigger. The figure shows
two striking features. Firstly, the associated yields with di-hadron
triggers are larger than with the single-hadron trigger. This is
because the di-hadron trigger selects more energetic jets than a
single hadron trigger. The second observation from the figure is that
the distributions are very similar in d+Au and Au+Au events,
indicating that the selected di-jets fragment like in the
vacuum. This also implies that there are no events with a larger
energy-difference between the two jets, as one might expect for Au+Au
events with energy loss. It should be realised, however, that the
fact that the \pt-selection for the two trigger hadrons are similar in
the current measurement is likely to select events with small
energy-asymmetry between the jets. 

The first model calculations for this kind of measurement are already
available \cite{Renk:2008km}. They indicate that by by changing the
\pt-cuts for the trigger hadrons independently, one can select events
with larger difference between the energies of the jets. The larger triggered data sample from
RHIC run-7 may allow to perform this measurement with more asymmetric
cuts.

\subsection{Multi-hadron cluster triggered correlations}
\begin{figure*}[ht]
\psfig{file=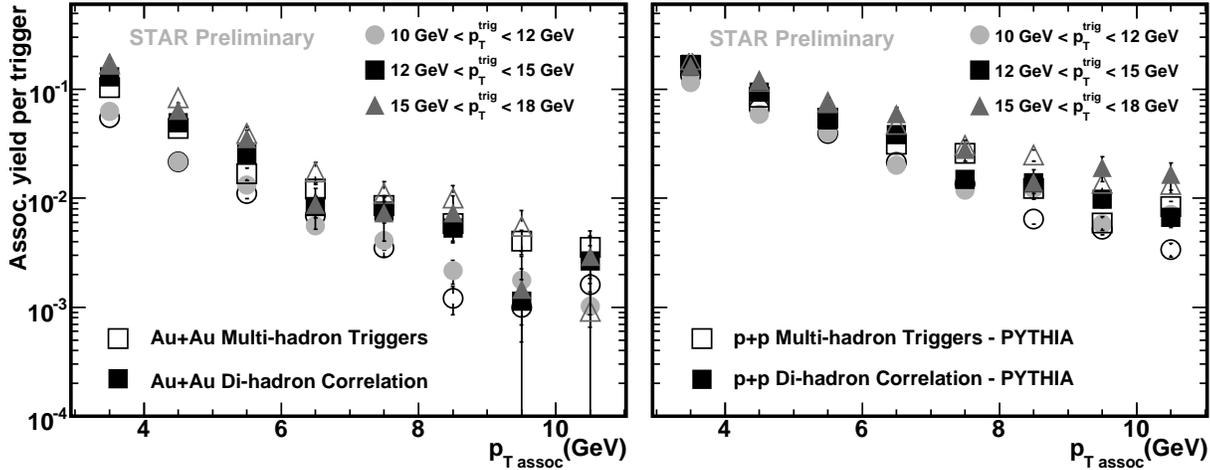,width=0.95\textwidth}
\caption{\label{fig:multi_had}Preliminary measurement of the recoil
  yield opposite multi-hadron trigger clusters consisting of at least
  one particle with $\pt > 5$  GeV/$c$ and possible additional particles
  with $\pt > 3$ GeV/$c$ inside a cone of $R= 0.3$ (open symbols) in
  0-12\% central Au+Au collisions. Results are shown
  for three selections of trigger-cluster energy and for data (left
  panel) and Pythia (right panel). Solid symbols show the results
  with a single-particle trigger for comparison.}
\end{figure*}

A different approach to reduce the jet-energy bias in di-hadron
correlation measurements is to cluster hadrons into a 'proto-jet'
and use this as the trigger object. A first analysis of this type is
being performed in STAR, using particles with $p_{T}^{seed} > 5$
GeV/$c$ as 'seed' particles. Secondary particles with $\pt>3$ GeV/$c$
are added to the cluster if they are within $R =
\sqrt{\deta^2+\dphi^2} < 0.3$ from the seed particle. The sum of the
transverse momenta of the particles in the cluster
is used as the multi-hadron trigger \pt. For details of the
analysis, see \cite{Haag:2008gi}. In particular, it should be noted
that at the moment, no correction is made for 'background triggers',
{\it i.e.}  random combinations of particles that form a trigger
cluster. The signal-to-background ratio for trigger clusters is
estimated to be 0.7 \cite{Haag:2008gi}.

Figure \ref{fig:multi_had} shows the recoil yield for such
multi-hadron cluster triggers in three different \pt{} ranges. The
left panel shows the measurement in 0-12\% central Au+Au collisions,
while the right panel shows the result for the same analysis on PYTHIA
events \cite{Sjostrand:2006za}. For comparison, results of the
di-hadron analysis (single particle trigger) are also shown (solid
symbols). It can be seen in the figure that the analysis with
multi-hadron cluster triggers gives similar results to the
single-hadron triggers, indicating that the underlying jet energy
selection is similar in both cases. A similar trend is seen in the
PYTHIA simulation. This implies that multi-hadron clusters with the
current cuts do not provide a significantly better measure of the jet
energy than the leading particle. Further studies with PYTHIA are
ongoing to determine whether for example including electromagnetic
energy in the cluster changes the result.

\section{Conclusion and outlook}
We have presented recent results on intermediate and high-\pt{}
multi-hadron correlation measurements. At intermediate \pt, two
results were highlighted that provide some insight in the origin of
the near-side associated yield at large \deta{}, the ridge. Both the
large baryon/meson ratios and the uniform event-by-event
\deta-structure indicate that the the ridge is likely
formed from bulk matter and not from jet fragments.

Measurements of the associated hadron distribution as a function of
the angle with the event plane show a clear evolution of the
correlation structure with path length, on both the near and the away
side. Comparisons to theoretical calculations are needed to interpret
the interplay between trigger bias and path length dependent energy loss.

At high \pt, STAR is exploring multi-hadron analyses to provide more
differential measures of energy loss and better constraints on parton
energy loss models. Back-to-back di-hadron triggered correlations may
provide a way to select events with large energy loss for further
analysis. The current data sample is not large enough to provide the
\pt-reach that is needed to fully exploit this technique. Another
analysis uses multi-hadron clusters as triggers to reduce the trigger
bias in correlation measurements. So far, the differences between the
multi-hadron triggered analysis and the analysis with single-hadron
triggers are found to be small. Simulated PYTHIA events are being used
to further explore the potential of this technique.

In the coming years, STAR will collect substantially larger
data samples for p+p and Au+Au events, which will allow more
differential analyses to further test jet quenching models. We
would like to urge theoretical physicists to help devise decisive
tests of our understanding of parton energy loss mechanisms in heavy ion
collisions.

\bibliographystyle{epj}
\bibliography{08HP_STARHighpt_mvanleeuwen}

\end{document}